\documentclass[onecolumn,superscriptaddress,aps,prb,preprint]{revtex4-1}

\usepackage{color}
\usepackage{hyperref}
\usepackage{amssymb}
\usepackage{amsmath}

\hypersetup{colorlinks,linkcolor=blue,citecolor=blue,urlcolor=blue}
 
\usepackage[version=4]{mhchem}
\usepackage{latexsym}
\usepackage[pdftex]{graphicx} %zum einbinden von xfig grafiken
\usepackage{xspace}

\begin{document}

\title{Lifting the spin-momentum locking in ultra-thin topological insulator films}

\author{Arthur Leis}
\affiliation{Peter Gr\"{u}nberg Institut (PGI-3), Forschungszentrum J\"{u}lich, 52425 J\"{u}lich, Germany}
\affiliation{J\"ulich Aachen Research Alliance (JARA), Fundamentals of Future Information Technology, 52425 J\"ulich, Germany}
\affiliation{Experimentalphysik IV A, RWTH Aachen University, Otto-Blumenthal-Stra\ss{}e, 52074 Aachen, Germany}

\author{Michael Schleenvoigt}
\affiliation{J\"ulich Aachen Research Alliance (JARA), Fundamentals of Future Information Technology, 52425 J\"ulich, Germany}
\affiliation{Peter Gr\"{u}nberg Institut (PGI-9), Forschungszentrum J\"{u}lich, 52425 J\"{u}lich, Germany}

\author{Vasily Cherepanov}
\affiliation{Peter Gr\"{u}nberg Institut (PGI-3), Forschungszentrum J\"{u}lich, 52425 J\"{u}lich, Germany}
\affiliation{J\"ulich Aachen Research Alliance (JARA), Fundamentals of Future Information Technology, 52425 J\"ulich, Germany}

\author{Felix L\"upke}
\affiliation{Peter Gr\"{u}nberg Institut (PGI-3), Forschungszentrum J\"{u}lich, 52425 J\"{u}lich, Germany}
\affiliation{J\"ulich Aachen Research Alliance (JARA), Fundamentals of Future Information Technology, 52425 J\"ulich, Germany}

\author{Peter Sch\"uffelgen}
\affiliation{J\"ulich Aachen Research Alliance (JARA), Fundamentals of Future Information Technology, 52425 J\"ulich, Germany}
\affiliation{Peter Gr\"{u}nberg Institut (PGI-9), Forschungszentrum J\"{u}lich, 52425 J\"{u}lich, Germany}

\author{Gregor Mussler}
\affiliation{J\"ulich Aachen Research Alliance (JARA), Fundamentals of Future Information Technology, 52425 J\"ulich, Germany}
\affiliation{Peter Gr\"{u}nberg Institut (PGI-9), Forschungszentrum J\"{u}lich, 52425 J\"{u}lich, Germany}

\author{Detlev Gr\"utzmacher}
\affiliation{J\"ulich Aachen Research Alliance (JARA), Fundamentals of Future Information Technology, 52425 J\"ulich, Germany}
\affiliation{Peter Gr\"{u}nberg Institut (PGI-9), Forschungszentrum J\"{u}lich, 52425 J\"{u}lich, Germany}

\author{Bert Voigtl\"ander}
\email{b.voigtlaender@fz-juelich.de}
\affiliation{Peter Gr\"{u}nberg Institut (PGI-3), Forschungszentrum J\"{u}lich, 52425 J\"{u}lich, Germany}
\affiliation{J\"ulich Aachen Research Alliance (JARA), Fundamentals of Future Information Technology, 52425 J\"ulich, Germany}
\affiliation{Experimentalphysik IV A, RWTH Aachen University, Otto-Blumenthal-Stra\ss{}e, 52074 Aachen, Germany}

\author{F. Stefan Tautz}
\affiliation{Peter Gr\"{u}nberg Institut (PGI-3), Forschungszentrum J\"{u}lich, 52425 J\"{u}lich, Germany}
\affiliation{J\"ulich Aachen Research Alliance (JARA), Fundamentals of Future Information Technology, 52425 J\"ulich, Germany}
\affiliation{Experimentalphysik IV A, RWTH Aachen University, Otto-Blumenthal-Stra\ss{}e, 52074 Aachen, Germany}
% \today

%\pacs{}
%\keywords{}

\maketitle

\textbf{Three-dimensional (3D) topological insulators (TIs) are known to carry 2D Dirac-like topological surface states in which spin-momentum locking prohibits back-scattering. %\cite{Chen2009, Zhang2009, Roushan2009}
When thinned down to a few nanometers, the hybridization between the topological surface states at the top and bottom surfaces results in a topological quantum phase transition, which can lead to the emergence of a quantum spin Hall phase. % \cite{Lu2010, Liu2010, Foerster2015, Foerster2016}.
%The latter is expected to display an insulating interior that is circumscribed by 1D conducting edge channels along the film perimeter.
Here, we study the thickness-dependent transport properties across the quantum phase transition on the example of (Bi$_{0.16}$Sb$_{0.84}$)$_2$Te$_3$ films, with a four-tip scanning tunnelling microscope.
Our findings reveal
%that reveal in detail how the insulating behavior emerges. While the conductivity of films above a critical thickness of 5 quintuple layers bears witness to transport in 2D topological surface states,
an exponential drop of the conductivity below the critical thickness. % is observed below this threshold.
The steepness of this drop indicates the presence of spin-conserving backscattering between the top and bottom surface states, effectively lifting the spin-momentum locking and resulting in the opening of a gap at the Dirac point.
Our experiments provide crucial steps towards the detection of quantum spin Hall states in %nanoscale 
transport measurements.
%, since a precise knowledge of the 2D conductivity in ultra-thin films is required to single out potential edge channels. 
}

\begin{figure}[b]
\includegraphics[width=\linewidth]{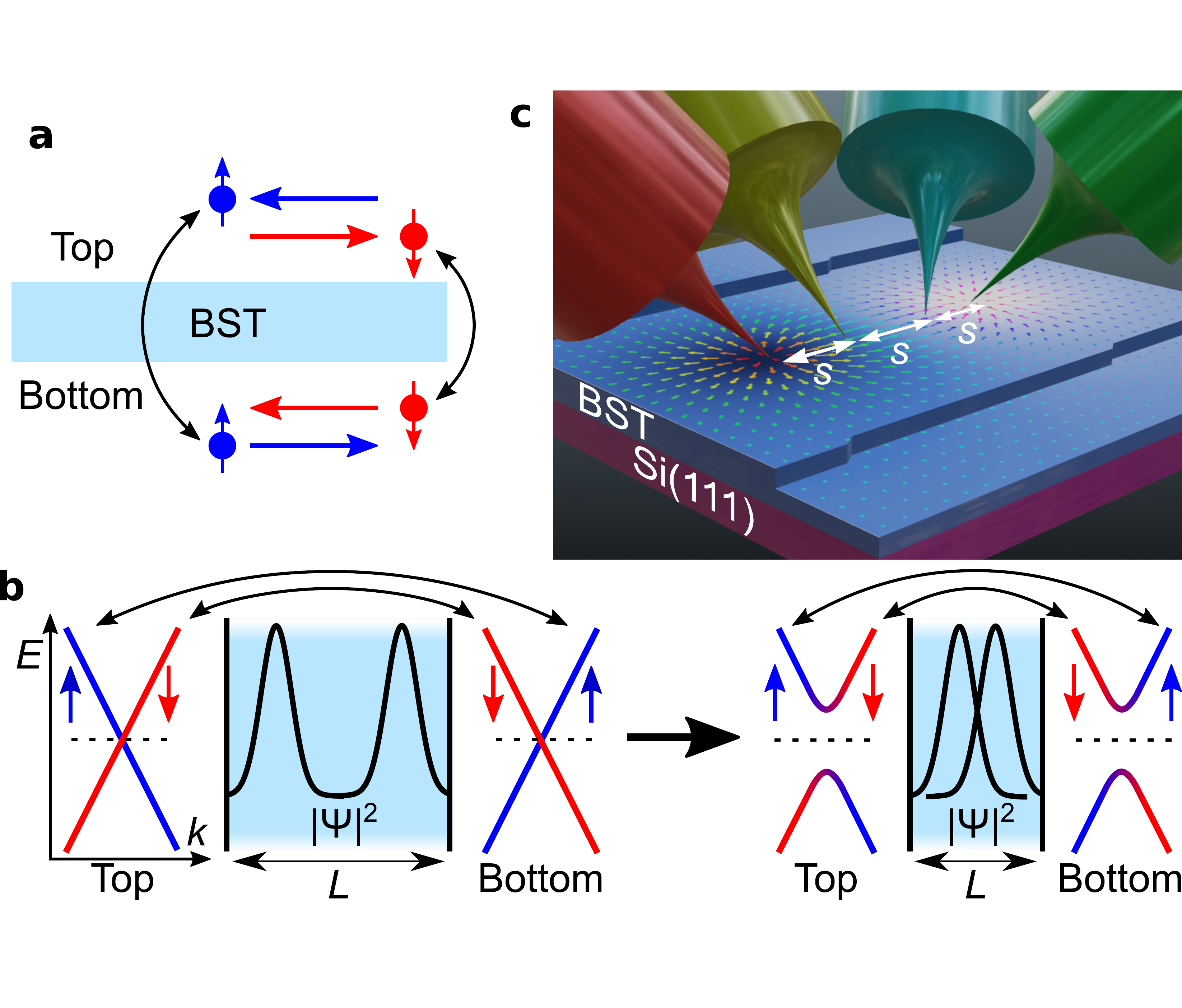}
\caption{\textbf{Scheme of the multi-tip STM experiment to detect scattering between topological surface states.}
% Sketch of the sample layout and measurement principle.
\textbf{a}, Scattering between topological surface states at the top and the bottom surfaces of a (Bi$_{1-x}$Sb$_x$)$_2$Te$_3$ thin film with thickness $L$ below a critical value of 5 QL. 
\textbf{b}, Because of the hybridisation of the overlapping wave functions of both topological surface states, a gap opens at the Dirac point. Hybridisation also enables spin-conserving scattering from $k$ to $-k$. 
\textbf{c}, Experimental setup. The boundary region of the BST film forms a wedge with step heights of single quintuple layers. 
%Over a width of $\sim 10 \, \mu$m, the TI layer thickness decreases from the maximum thickness of $12 \,$QL down to the Si substrate in single quintuple-layer steps orthogonally to the film boundary. 
Positioning the tips of a four-tip STM on a single terrace allows measuring the sheet conductivity $\sigma$ as a function of the film thickness $L$.}
\label{fig:fig1}
\end{figure}

Since the discovery of 3D topological insulators
%,with band gaps of up to $300\,$meV
\cite{Zhang2009, Xia2009, Chen2009, Hsieh2009}, an increasing number of novel topological phases have been realized. %based on this material system.
For example, in magnetically doped 3D TI thin films a quantum anomalous Hall (QAH) phase with 1D chiral edge states was reported  %(Bi$_{1-x}$Sb$_x$)$_2$Te$_3$ 
 \cite{Chang2013, Kou2014, Chang2015}. 
A more recent example are coexisting QAH and axion insulator phases in a stoichiometric magnetic topological insulator
% MnBi$_2$Te$_4$ 
\cite{Deng2020, Liu2020}.
The emergence of these exotic topological phases is underpinned by the breaking of time-reversal symmetry, which opens a gap at the Dirac point and leads to massive Dirac fermions. 

Yet another possibility to gap out the Dirac point of a 3D TI is to reduce its thickness below a critical value at which the topological surface states (TSS) at the top and bottom of the film start to interact. 
In the prototypical 3D TI (Bi$_{1-x}$Sb$_x$)$_2$Te$_3$ (BST), this occurs at thicknesses below $\sim5$ quintuple layers (QL). 
While $180^\circ$-backscattering of electrons in the TSS is prohibited by spin-momentum locking as long as the states on opposite surfaces are strictly separated \cite{Roushan2009}, spin-conserving backscattering becomes possible if opposite TSS interact and electrons are able to scatter from the top to the bottom surface and vice versa, as illustrated in Fig.~\ref{fig:fig1}a. In such ultra-thin TI films, a quantum spin Hall (QSH) phase with one-dimensional helical edge states is predicted to emerge \cite{Lu2010, Liu2010, Foerster2015, Foerster2016}. Intuitively, the latter can be understood as a remnant of the 2D TSS on the side faces of the film, when upon film thinning they are reduced to 1D edges.

Fig.~\ref{fig:fig1}b indicates that the coupling between opposite TSS in an ultra-thin TI film has two consequences: On the one hand, the band structure will change, from two separate Dirac cones to a single gapped structure. Note that the continuity of transport in a loop around the material means that the Dirac cones on the top and bottom surfaces have interchanged spins (Fig.~\ref{fig:fig1}b, left). Precisely, this allows spin-conserving scattering from $k$ to $-k$ if top and bottom surface states interact (Fig.~\ref{fig:fig1}b, right), which effectively corresponds to a lifting of spin-momentum locking in the TSS. Therefore, as a second consequence in addition to the modification of their dispersion, the lifetime of the electrons in the surface states is expected to drop. While the gap opening at the Dirac point has been detected in angle-resolved photoemission spectroscopy and scanning tunnelling spectroscopy experiments on ultra-thin TI films \cite{Zhang2010, Jiang2012, Stroscio2013}, the second effect has yet not been observed, since it requires the systematic measurement of the transport properties of pristine samples, which is difficult in lithographically patterned samples, because the processing tends to degrade the ultra-thin films. For this reason, we apply here the methodology of multi-tip scanning tunnelling microscopy (STM) as a ``multimeter on the nanoscale'' to study (Bi$_{1-x}$Sb$_x$)$_2$Te$_3$ films \textit{in situ} \cite{Bauer2016, Luepke2017, Fukui2020}.

\begin{figure}[b]
\includegraphics[width=0.7\linewidth]{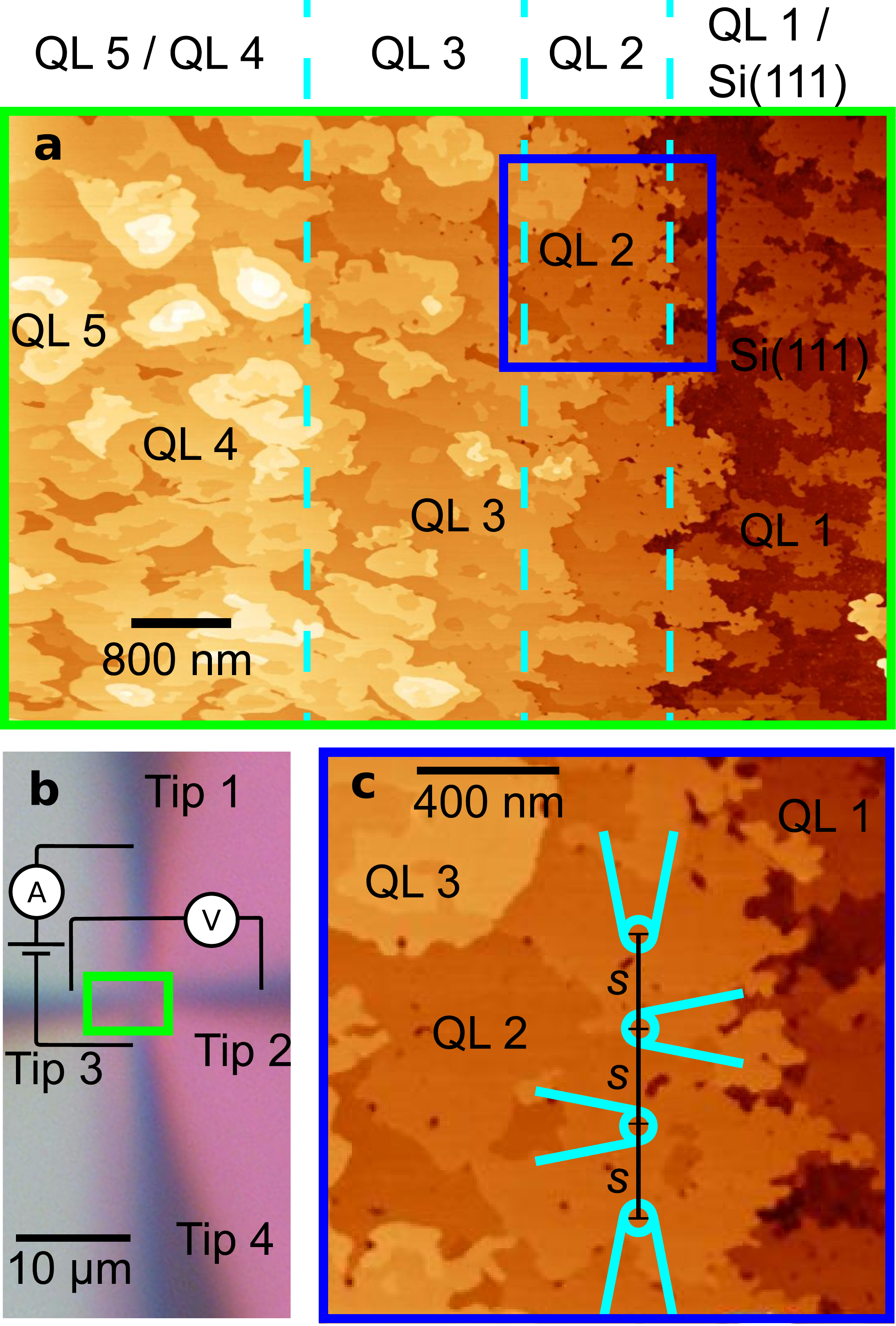}
\caption{\textbf{Measurement configuration and tip positioning method.}
\textbf{a}, A large overview STM scan is performed with one of the tips to map the topography of the TI film boundary region, as indicated by the green rectangle. This large overview scan serves as a reference map to place all four tips on a single terrace. In the entire overview area, single steps of QL-height are seen, revealing the wedge-shaped structure of the film boundary region.
\textbf{b}, Optical microscope image of the final tip configuration.
\textbf{c}, The tips are navigated to the desired positions using a method of overlapping STM scans\cite{Leis2020}. The tip positions are located by identifying the same topographic features in the small scans and in the overview scan.}
\label{fig:fig2}
\end{figure}

Samples are prepared by molecular beam epitaxy (MBE) with a shadow mask, which allows the deposition of a (Bi$_{1-x}$Sb$_x$)$_2$Te$_3$ wedge in which the film thickness increases in steps of single quintuple layers from 1 $\mathrm{QL}\approx 1 \,$nm at the edge of the film to 12 QL in its centre (Fig.~\ref{fig:fig1}c, see Methods for details). We have chosen a stoichiometry of $x = 0.84$, because for this value the Fermi level in the bulk of the film is located in the bulk band gap close to the Dirac point, which reduces parasitic charge transport through the interior of the film \cite{Luepke2018, Just2020}. Large-scale STM scans at the edge of the film show the step-wise increase of the film thickness (Fig.~\ref{fig:fig2}a) from the Si(111) substrate. For the transport measurements, the four tips are individually navigated into the boundary region of the film, as monitored with an optical microscope (Fig.~\ref{fig:fig2}b). To measure the local conductivity of the TI film, the four STM tips are positioned in a linear configuration on a single terrace, with a  distances of $s \approx 250 \,$nm between adjacent tips (Fig.~\ref{fig:fig2}c). In this configuration, the four point resistance is measured, from which the 2D sheet conductivity $\sigma$ can be calculated (see Methods for details). Topography scans demonstrate that the film surface, including the local terrace structure, is still intact after the electrical measurements. Thus, our experiments yield sheet conductivities for well-defined film thicknesses $L$. 

\begin{figure*}
\includegraphics[width=\textwidth]{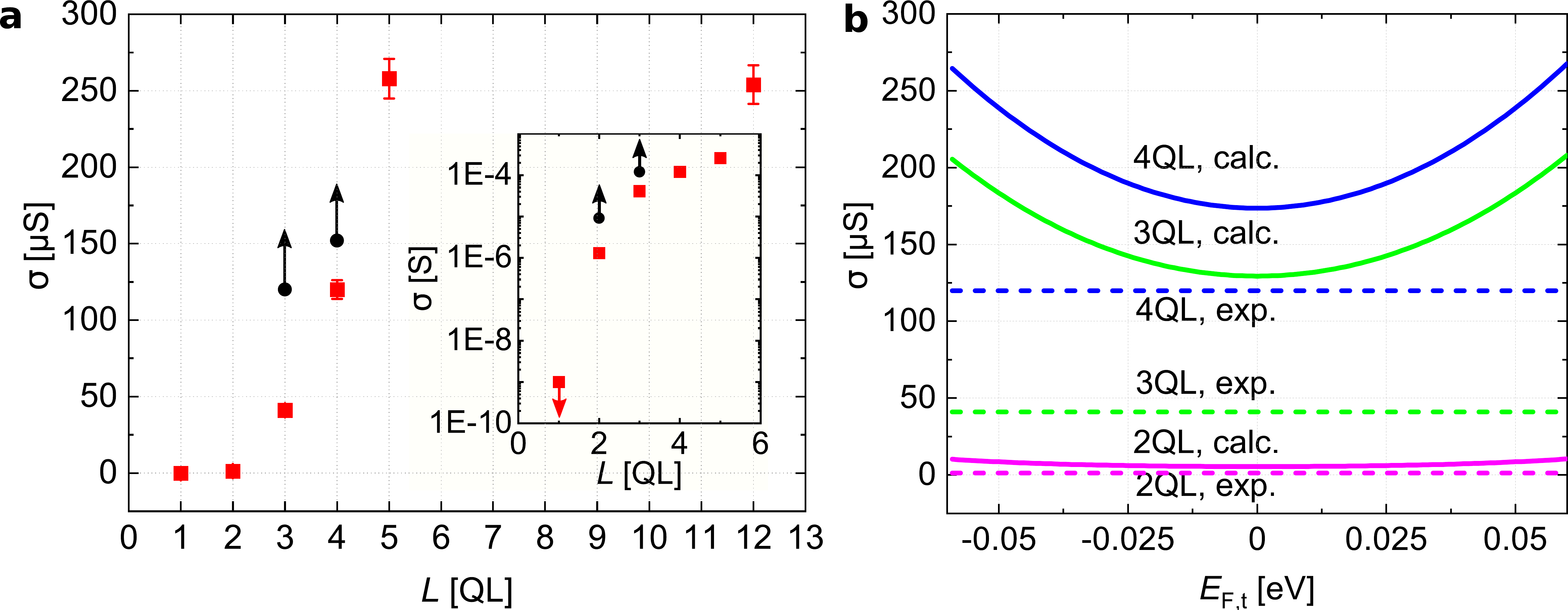}
\caption{\textbf{Measured thickness-dependent sheet conductivity compared to calculations based on the TSS band structure.}
\textbf{a}, Thickness-dependent conductivity (red squares with error bars) as obtained from four-point measurements on single terraces of different layer thickness in the TI film. The minimum of the calculated conductivity, obtained when varying $E_\mathrm{F,t}$, $E_\mathrm{F,b} \in [-50\,\mathrm{meV};50\,\mathrm{meV}]$ and $r_\mu \in [0.3;1]$, is indicated by black circles. The inset shows a logarithmically scaled plot of the same data. 
\textbf{b}, Calculated conductivity as function of $E_\mathrm{F,t}$, with $r_\mu = 1$ and $E_\mathrm{F,b} = -50\,$meV
(Eq. \ref{eq:cond}).}
\label{fig:fig3}
\end{figure*}

The sheet conductivity displays an exponential increase from $1 \,$QL to $5 \,$QL (Fig.~\ref{fig:fig3}a). Note that for $L> 5\,$QL it is not possible to realize the four-point measurement on a single terrace, because the terraces are too narrow to reliably place the four tips on them. However, large-scale conductivity measurements far away from the film edge with a tip spacing of $s=50\,\mu$m yield, within measurement error, the same conductivity for the $12 \,$QL interior of the film as for the $5\,$QL terrace. This saturation of the thickness-dependent conductivity verifies that the TSS at the top and bottom of the film dominate the charge transport. A possible parasitic contribution from bulk states would result in a linear dependence of the conductivity on the film thickness $L$. The sheet conductivity $\sigma$ is thus given by carrier concentrations $n$ and carrier mobilities $\mu$ in the top and bottom TSS (Drude model),
\begin{equation}
\sigma = e \left[ \mu_\mathrm{t} n_\mathrm{t}(\Delta,E_\mathrm{F,t}) + \mu_\mathrm{b} n_\mathrm{b}(\Delta,E_\mathrm{F,b}) \right],
\label{eq:cond}
\end{equation}
where $e$ is the elementary charge. Apart from the Fermi energy $E_\mathrm{F}$, which we consider relative to the Dirac point \cite{Just2020} and which -- due to the presence of the substrate -- may differ between the top and bottom surfaces of the film, it is the gap $\Delta$ in the topological surface state at the Dirac point that predominantly determines $n_\mathrm{t}$ and $n_\mathrm{b}$, through the dispersion relation \cite{Skinner2013, Lu2010}  
\begin{equation}
E(k,L) = \pm \sqrt{[ \hbar k v_\mathrm{F} ]^2 + [\Delta(L) /2 ]^2}
\label{eq:gap_dispersion}
\end{equation}
of the massive Dirac fermions in the vicinity of the $\Gamma$-point. $\Delta(L)$ has been measured spectroscopically on Sb$_2$Te$_3$ \cite{Jiang2012}, yielding the opening of a gap below $5\,$QL that increases up to a value of $\Delta\simeq250\,$meV for $2\,$QL, in agreement with theoretical predictions \cite{Foerster2016}. Since our sample has similar composition, we use these values for $\Delta(L)$. $v_\mathrm{F}$ is the Fermi velocity, which determines the slope of the Dirac cone and therefore does not depend on the film thickness. We use a value of  $v_\mathrm{F}=4.2 \times 10^5 \,$m/s, as measured in a previous photoemission study on samples of identical composition \cite{Kellner2015} and confirmed by an interpolation (see Methods). Before we can determine $\sigma$, we still need the Fermi levels. While $E_\mathrm{F,t}$ is accessible through photoemission experiments,  $E_\mathrm{F,b}$ follows from gate-dependent transport experiments (see Supplementary Note~2). For the present sample, we find $E_\mathrm{F,t}\simeq50$\,meV and $E_\mathrm{F,b}\simeq-50$\,meV for $L\geq5$\,QL (see Methods). 

With these parameters, we can calculate the expected sheet conductivities for different $L$ from Eq.~\eqref{eq:cond} under the assumption that $n_\mathrm{t}$ and $n_\mathrm{b}$ change due to the gap opening, according to 
  \begin{equation}
\frac{\mu_\mathrm{t}(L)}{\mu_\mathrm{t}(L')} = \frac{m^*_\mathrm{t}(L')}{m^*_\mathrm{t}(L)}= v_\mathrm{F}^2 	\frac{m^*_\mathrm{t}(L')}{\sqrt{{E_\mathrm{F,t}}^2 + [\Delta(L)/2]^2}} 	,
\label{eq:mu0exp}
\end{equation}
where $m^{\ast}(L)$ is the thickness-dependent effective carrier mass (see Methods). Here, the electron mobility $\mu_\mathrm{t}=e\tau_0/m^*$ at the top of the film changes only through the change in the band structure associated with the gap opening and the resulting change in effective mass $m^*$, while the scattering time $\tau_0$ stays fixed.
In addition, we assume a ratio $r_\mu$ between the top an bottom mobilities. The minimum sheet conductivities $\sigma(L)$ predicted by this model (see Methods for details) for $L'=5\,$QL are displayed as circles in Fig.~\ref{fig:fig3}a. Evidently, this model cannot explain the sharp drop of the experimentally determined sheet conductivity, even if we allow for a variation of the parameter $E_\mathrm{F,t}$ from its experimentally confirmed value at $L\geq5$\,QL (Fig.~\ref{fig:fig3}b) and for a variation of $r_\mu$ and $E_\mathrm{F,b}$ within reasonable ranges. We therefore conclude that the measured thickness-dependent conductivity drops too sharply to be explained by a mere band structure effect.
 
\begin{figure}
\includegraphics[width=\linewidth]{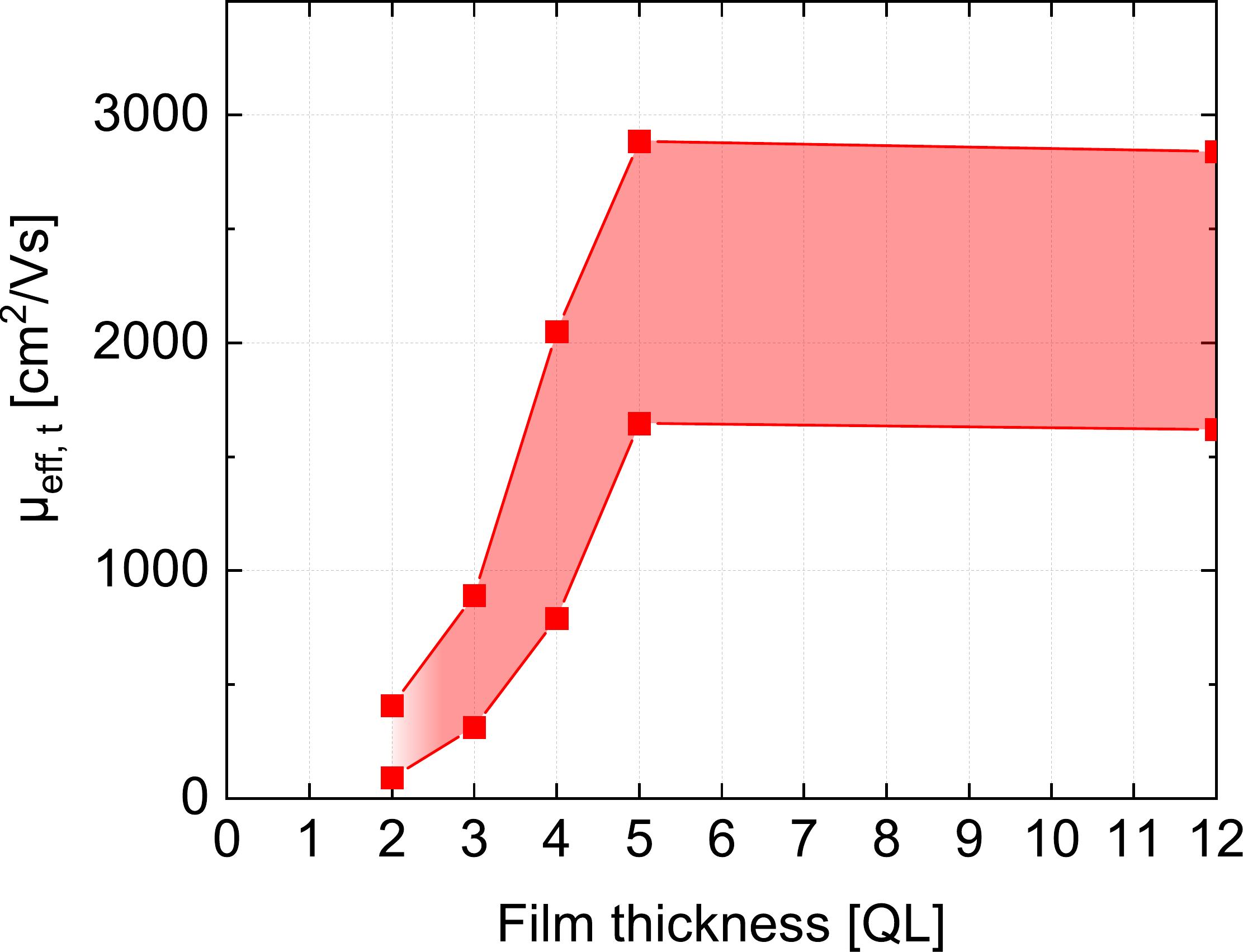}
\caption{\textbf{Thickness-dependent electron mobility as inferred from experimental results.} For $L\leq4\,$QL, the mobility $\mu_\mathrm{eff,t}$ is the mobility in the coupled top and bottom TSS. It is reduced below the value for $L\geq5\,$QL by the lifting of the spin-momentum locking in ultra-thin films, as effected by the onset of of spin-conserving $180^\circ$-backscattering between the top and the bottom surfaces of the film (inter-TSS scattering). For $L\geq5\,$QL, the mobility $\mu_\mathrm{eff,t}$ corresponds to the electron mobility in the top TSS, limited by intra-TSS scattering only.
The band indicated in red represents the range of mobility values obtained for the range of the parameters $E_\mathrm{F,t}$, $E_\mathrm{F,b}$ and $r_\mu$ discussed in the main text and the Methods section. The corresponding inter-TSS scattering times are
 $11 \,$fs $< \Tilde{\tau}_\mathrm{4QL} <56 \,$fs, $10 \,$fs $< \Tilde{\tau}_\mathrm{3QL} < 19 \,$fs and $8 \,$fs $< \Tilde{\tau}_\mathrm{2QL} < 45 \,$fs.
For $L\geq5\,$QL, $E_\mathrm{F,t}$ is known and the red band is influenced only by $r_\mu$ and $E_\mathrm{F,b}$.}
\label{fig:fig4}
\end{figure}
 
If a new efficient scattering channel appears at small film thicknesses $L<5$\,QL, Matthiesen's rule 
$\tau^{-1} =\tau_0^{-1} + \Tilde{\tau}^{-1}$ predicts a decrease in the overall scattering time $\tau$, which could explain the sharp drop in the sheet conductivity by a corresponding drop in the mobility. In fact, once the top and bottom surface states interact, we expect the emergence of \textit{inter-TSS} scattering $\Tilde{\tau}$, in addition to \textit{intra-TSS} scattering $\tau_0$ which originates from surface (or interface) defects and the electron-phonon interaction.  
Including this additional scattering mechanism in our model, we obtain from the experimental data in Fig.~\ref{fig:fig3}a the effective electron mobility in the top TSS according to Matthiesen's rule
\begin{equation}
\mu_\mathrm{eff,t}(L) = \frac{\mu_\mathrm{t}(L)\Tilde{\mu}_\mathrm{t}(L)}{\mu_\mathrm{t}(L) + \Tilde{\mu}_\mathrm{t}(L)} ,
\label{eq:mu_eff_t}
\end{equation}
being composed of the intra-TSS mobility $\mu_\mathrm{t}$ and the inter-TSS mobility $\Tilde{\mu}_\mathrm{t}$, as a function of film thickness $L$ (Fig.~\ref{fig:fig4}). Because of some uncertainty in the parameters $E_\mathrm{F,t}$,  $E_\mathrm{F,b}$, and $r_\mu$ at the corresponding film thickness, it is only possible to display a band within which the mobility must lie. Nevertheless, the plot shows a dramatic decrease of the total mobility of the top surface $\mu_\mathrm{eff,t}$. This decrease explains the deviation between experimental and calculated data points in Fig.~\ref{fig:fig3}a. Regarding the strength of the inter-TSS scattering, we find that $\Tilde{\tau}<\tau_0$ for all thicknesses $2\,\mathrm{QL}\leq L \leq4\,\mathrm{QL}$, i.e. the inter-layer scattering dominates the intra-layer scattering. For 2\,QL, $\tau_0$ is between 2 and 8 times larger than $\Tilde{\tau}$, and values for $\Tilde{\tau}$ range between approximately $10$ and $50$\,fs. We also find that the dependence of $\Tilde{\tau}$ on $L$ is weak, as $11 \,$fs $< \Tilde{\tau}_\mathrm{4QL} <56 \,$fs, $10 \,$fs $< \Tilde{\tau}_\mathrm{3QL} < 19 \,$fs and $8 \,$fs $< \Tilde{\tau}_\mathrm{2QL} < 45 \,$fs. This can be rationalized as a compound effect of an increasing matrix element for scattering as a result of rising wave function overlap on the one hand and a decreasing density of states at the Fermi level to scatter into as the band gap opens on the other. 
%The scattering time $\tau_\mathrm{x}$ will be small, because the interaction of top and bottom surface states effective lifts the spin-momentum locking and spin-conserving backscattering becomes possible by changing from the top to the bottom TSS and vice versa, thus opening an efficient decay channel for the current in the film (Fig.~\ref{fig:fig1}a). 

In conclusion, we find that in ultra-thin films of topological insulators the scattering between topological surface states on opposite faces of the film significantly reduces the sheet conductivity beyond what is expected from the opening of a Dirac gap alone. This inter-TSS scattering is in fact the dominant scattering mechanism, because the interaction of top and bottom surface states effectively lifts the spin-momentum locking and spin-conserving $180^\circ$-backscattering becomes possible between the top to the bottom surface state, thus opening an additional scattering channel (Fig.~\ref{fig:fig1}a).
Even if the Fermi level is not located in the Dirac surface gap and the influence of the gap on the transport is expected to be small, the additional inter-TSS scattering decreases the surface state conductivity. This effect helps to establish an insulating 2D interior against which 1D edge states, e.g. in a QSH phase, prevail.
% thereby supporting the insulating 2D interior against which conductive 1D edge states, e.g. in a QSH phase, prevail.
The measurement of charge transport with multi-tip STM on pristine surfaces on the sub-$\mu$m scale is thus an important step towards the detection and characterization of possible 1D edge states.

\section*{Methods}

\subsection*{Sample preparation}

The (Bi$_{0.16}$Sb$_{0.84}$)$_2$Te$_3$ thin films were grown on a silicon-on-insulator (SOI) substrate, the latter consisting of a degenerately doped Si(100) handle wafer, a $300\,$nm oxide layer, and an undoped $70\,$nm Si(111) template layer. Employing a thin intrinsic Si layer as template for growth reduces the substrate sheet conductivity in our experiments to $\sim2 \,$nS.

The substrate is cleaned by Piranha solution (H$_2$SO$_4$:H$_2$O$_2$ 2:1) and a HF (1\%) dip to remove organic contaminations and native oxides, respectively, while also supplying a protective hydrogen passivation. After transfer into the MBE chamber (base pressure $p \simeq 1 \times 10^{-9}$\,mbar), the substrate is heated to 700\,${}^\circ$C for 10 minutes for hydrogen desorption and subsequently cooled down to 262\,${}^\circ$C for growth. Bi, Sb and Te are evaporated from standard Knudsen effusion cells at temperatures of 440\,${}^\circ$C, 470\,${}^\circ$C and 330\,${}^\circ$C, respectively, resulting in a flux ratio of 1:20:120. Initially, the substrate is flushed with Te to saturate the Si dangling bonds before supplying Bi and Sb for TI growth. 

Growth takes place through a removable shadow mask in order to achieve a TI film with boundaries on the SOI substrate without the need of ex situ processing. In the boundary region of the shadow mask, the thickness of the TI film decreases in steps of single quintuple layers from the maximum film thickness of 12\,QL to zero, exposing the Si(111) template layer.

\subsection*{Four-point transport measurements}
After growth, the sample is transferred in vacuum ($p\leq1\times10^{-9}\rm\,mbar$) into the room-temperature four-tip STM (base pressure $p\simeq4\times10^{-10}\rm\,mbar$). The vacuum transfer of the sample and the absence of any processing steps after the growth enable us to measure transport properties of the pristine TI film, avoiding any detrimental influence of passivation or lithography steps. 

The STM tips are electrochemically etched tungsten wires. We use an optical microscope to position four STM tips in the boundary region of the (Bi$_{0.16}$Sb$_{0.84}$)$_2$Te$_3$ film on the sample (coarse navigation). In the optical microscope, the wedge-shaped boundary region appears as a blend of colors between grey ($12 \, $QL TI film) and pink (Si substrate) (Fig. \ref{fig:fig2}c). 

After optical coarse positioning, large-scale STM images are recorded to obtain an overview of the topography of the TI film in the corresponding area. Then the STM tips are individually and precisely positioned on a chosen terrace of specific thickness $L$ (which can be determined by counting step edges) in a linear four-point configuration with a distance $s \approx 250$\,nm between the tips. To this end, we employ a positioning method that relies on overlapping small-area STM scans, performed successively with each tip \cite{Leis2020}. 

Once the four tips have been positioned laterally, they are driven from the tunnelling regime into contact with the sample surface. In this electrical point contact regime, the sheet conductivity $\sigma$ of the TI film is measured by repeatedly recording the voltage drop $V$ between the two inner tips as a function of the current $I$ injected between the two outer tips. For each $L$, experiments are repeated in at least two different areas of the sample (see Supplementary Note 1). In the given contact geometry, the sheet conductivity $\sigma$ of a 2D film is given by \cite{Voigtlaender2018}
\begin{equation*}
\sigma= \frac{\ln 2}{\pi}\frac{I}{V}.
\end{equation*}
With an inter-tip distance $s$ smaller than the distance between the tips and the nearest step edge, conductivity contributions from neighbouring terraces are negligible\cite{Miccoli2015}. 
We note that the conductivity of $1 \,$nS measured for $L=1 \,$QL is an upper boundary, since the conductivity of the bare Si substrate is of the same order of magnitude ($\sim2 \,$nS). 

After completion of the electrical measurements, the tips are retracted and the sample area is imaged again with STM. In these images, the contact points of the tips are usually discernible as small spots of typically only $\sim 1 \,$nm height (see Supplementary Fig. 1).

\subsection*{Sample characterization}

 The composition of our (Bi$_{1-x}$Sb$_x$)$_2$Te$_3$ thin film is determined ex situ by means of Rutherford backscattering spectroscopy after the electrical four-point experiments. From these measurements, we find that the Sb concentration in the sample is $x = 0.84$.

With the exact material composition known, we determine the position of the Fermi level $E_\mathrm{F,t}$ with respect to the Dirac point as well as the Fermi velocity $v_{\rm F}$. To this end, we interpolate spectroscopic data of samples grown previously in the same MBE system. Both parameters are approximately linearly dependent on the Sb concentration between $x = 0.5$ and $x = 1$ \cite{Zhang2011}. In a previous angle-resolved photoemission study, the Fermi level $E_\mathrm{F,t}$ for a (Bi$_{1-x}$Sb$_x$)$_2$Te$_3$ thin film with $x = 0.94$ was found to be located at the Dirac point, i.e. $E_\mathrm{F,t} \approx 0$  \cite{Kellner2015}. The corresponding Fermi velocity was determined as $v_\mathrm{F} \approx 3.8 \times 10^5 \,$m/s. Another sample grown in the same system with a Sb concentration of $x = 0.47$ had $E_\mathrm{F,t} \approx 250 \,$meV and $v_\mathrm{F} \approx 5.6 \times 10^5 \,$m/s \cite{Luepke2018}. Using these two samples as a reference, we obtain the $E_\mathrm{F,t} \approx 50 \,$meV and $v_\mathrm{F} \approx 4.2 \times 10^5 \,$m/s for the present (Bi$_{0.16}$Sb$_{0.84}$)$_2$Te$_3$ sample.

\subsection*{Calculation of the charge carrier density}

The dispersion relation Eq.~\ref{eq:gap_dispersion} in the vicinity of the Dirac point corresponds to a density of states of 
\begin{equation*}
D(\Delta(L),E) =	\frac{\left|E\right|}{2 \pi \left( \hbar v_\mathrm{F} \right)^2}  
\Theta(\left|E\right| - \Delta(L)/2)   ,
\label{eq:DOS}
\end{equation*}
with the energy measured $E$ relative to the Dirac point and $\Theta$ being the Heaviside step function \cite{Skinner2013}.
From the density of states, we calculate the charge carrier density as
\begin{equation*}
n(\Delta(L),E_\mathrm{F}) = \int_0^\infty D(\Delta(L),E) \left[ f_\mathrm{n}(E,E_\mathrm{F}) + f_\mathrm{p}(E,E_\mathrm{F}) \right]	\mathrm{d}E	,
\label{eq:ntot}
\end{equation*}
with the Fermi distribution function
\begin{equation*}
f_\mathrm{n}(E,E_\mathrm{F}) = f_\mathrm{p}(E,-E_\mathrm{F}) = \left( 1 + \mathrm{exp}\left( \frac{E-E_\mathrm{F}}{k_\mathrm{B}T} \right) \right)^{-1}
\end{equation*}
for electrons (n) and holes (p). 
The charge carrier density $n$ enters the Drude expression in Eq.~\ref{eq:cond}.
The TSS gaps $\Delta(\mathrm{2\,QL}) \approx 250\,$meV, $\Delta(\mathrm{3\,QL}) \approx 60 \,$meV, $\Delta(\mathrm{4\,QL}) \approx 25 \,$meV are taken from Ref. \onlinecite{Jiang2012} and were measured on pure Sb$_2$Te$_3$. These results are in agreement with theoretical predictions \cite{Foerster2016}.\\

\subsection*{Modelling the surface conductivity: no inter-TSS scattering}

The mobility in the TSS at the top surface of the film is given by 
\begin{equation*}
\mu_\mathrm{t}(L) = \frac{e}{m^\ast(L)} \tau_0,
\label{eq:mu0}
\end{equation*}
where $\tau_0$ is the intra-TSS scattering time due to surface defects and electron-phonon scattering. Because of spin-momentum locking in the TSS, spin-conserving $180^\circ$-backscattering events do not contribute to $\tau_0$. Also, $\tau_0$ does not depend on the film thickness $L$, since the primary sources of scattering are not influenced by the physical distance between the two interfaces film/vacuum and film/substrate. In contrast, the effective mass $m^\ast$ at the Fermi wave vector $k_\mathrm{F}$, being a property of the band structure, is affected by the presence of a gap $\Delta(L)$ at the Dirac point, and therefore for $L<5\,\mathrm{QL}$ depends on $L$.
Hence, 
\begin{equation*}
\frac{\mu_\mathrm{t}(L)}{\mu_\mathrm{t}(L')}=\frac{m^\ast(L')}{m^\ast(L)}
\end{equation*}
holds.
With Eq.~\ref{eq:gap_dispersion}, the effective mass is given by \cite{Ariel2012} %\cite{Zou2011}
\begin{align*}
m^\ast(L) &= \hbar^2 k \left( \frac{\mathrm{d}E}{\mathrm{d}k} \right)^{-1} \Bigr|_{k = k_\mathrm{F,t}}\\
&= \hbar^2 k \left( 2k (\hbar v_\mathrm{F})^2 \frac{1}{2 \sqrt{[\hbar v_\mathrm{F} k]^2 +[\Delta(L)/2]^2}} \right)^{-1} \Bigr|_{k = k_\mathrm{F,t}}\\
&= \frac{1}{v_\mathrm{F}^2} \sqrt{{E_\mathrm{F,t}}^2 + [\Delta(L)/2]^2}.
\label{eq:meff}
\end{align*}
We note the parametric dependence on $E_\mathrm{F,t}$, which in turn may depend on $L$. Eq.~\ref{eq:mu0exp} follows directly from this equation. Setting $L'= 5$\,QL in Eq.~\ref{eq:mu0exp}, we calculate $\mu_\mathrm{t}(L)$ for any $L<5$\,QL in terms of $\mu_\mathrm{t}(5\,\mathrm{QL})$, which in turn is given by (Eq.~\ref{eq:cond})
\begin{equation*}
\mu_\mathrm{t}(5\,\mathrm{QL})=\frac{\sigma (5\,\mathrm{QL})}{e[n_\mathrm{t}(5\,\mathrm{QL}) + r_\mu n_\mathrm{b}(5\,\mathrm{QL})]},
\end{equation*}
where we have assumed $\mu_\mathrm{b}(L) = r_\mu  \mu_\mathrm{t}(L)$, i.e.~a ratio between the mobilities in the top and bottom TSS. The resulting expression is 
\begin{equation*}
\mu_\mathrm{t}(L)= v_\mathrm{F}^2\frac{\sigma (5\,\mathrm{QL}) \ m^\ast(5\,\mathrm{QL}, E_\mathrm{F,t}) }{e[n_\mathrm{t}(5\,\mathrm{QL}) + r_\mu n_\mathrm{b}(5\,\mathrm{QL})]}\frac{1}{\sqrt{{E_\mathrm{F,t}}^2 + [\Delta(L)/2]^2}}
\end{equation*}
which can be inserted into Eq.~\ref{eq:cond}
\begin{equation*}
\sigma(L) = e \mu_\mathrm{t}(L) [n_\mathrm{t}(L) + r_\mu  n_\mathrm{b}(L)]
\end{equation*}
to calculate the sheet conductivity as a function of film thickness, only considering intra-TSS scattering (Fig.~\ref{fig:fig3}b). We vary the parameters  ${E_\mathrm{F,t}}$, ${E_\mathrm{F,b}}$ and $r_\mu$ in this equation in the ranges $[-50\,\mathrm{meV};50\,\mathrm{meV}]$ and $[0.3;1]$, respectively, always finding sheet conductivities that are too large in comparison with the experiment.  

\subsection*{Modelling the surface conductivity: including inter-TSS scattering}

The scattering between the top and bottom TSS is governed by Fermi's golden rule, i.e.
\begin{equation*}
\frac{1}{\Tilde{\tau}_\mathrm{t}} = \frac{2 \pi}{\hbar} D(E_\mathrm{F,t}) {|V_\mathrm{t \rightarrow b}|}^2,
\label{eq:fermi}
\end{equation*}
with an equivalent equation for $\Tilde{\tau}_\mathrm{b}$. Evidently, the modulus-squared matrix elements for the scattering from the top to the bottom ($|V_\mathrm{t \rightarrow b}|^2$) and from the bottom to the top ($|V_\mathrm{b \rightarrow t}|^2$) surfaces must be the same. The contribution of inter-TSS scattering to the mobility is given by 
\begin{equation*}
\Tilde{\mu_\mathrm{t}}(L) = \frac{e}{m^\ast(L, E_\mathrm{F,t})} \Tilde{\tau}_\mathrm{t}, 
\label{eq:mu_inter}
\end{equation*}
again with an equivalent expression for $\Tilde{\mu_\mathrm{b}}$. Together, both expressions yield
\begin{align*}
\Tilde{\mu}_\mathrm{b} &= \Tilde{\mu}_\mathrm{t} \, \frac{\Tilde{\tau}_\mathrm{b}}{\Tilde{\tau}_\mathrm{t}}\, \frac{m^\ast(L, E_\mathrm{F,t})}{m^\ast(L, E_\mathrm{F,b})}\\
&= \Tilde{\mu}_\mathrm{t} \, \frac{D(E_\mathrm{F,t}) \sqrt{{E_\mathrm{F,t}}^2 + {[\Delta(L)/2]}^2}}{D(E_\mathrm{F,b}) \sqrt{{E_\mathrm{F,b}}^2 + {[\Delta(L)/2]}^2}}.			
\end{align*}
Because of the increasing wave function overlap between the two TSS, we expect that in thermodynamic equilibrium $E_\mathrm{F,t}$ approaches $E_\mathrm{F,b}$ as $L$ decreases, such that because of Fermi's golden rule  $\Tilde{\tau}_\mathrm{t}\rightarrow \Tilde{\tau}_\mathrm{b}$, and finally also  $\Tilde{\mu}_\mathrm{t}\rightarrow \Tilde{\mu}_\mathrm{b}$. In our modeling, we assume $E_\mathrm{F,t}=E_\mathrm{F,b}$, $\Tilde{\tau}_\mathrm{t}= \Tilde{\tau}_\mathrm{b}$ and $\Tilde{\mu}_\mathrm{t}= \Tilde{\mu}_\mathrm{b}$ for 
all $L \leq 4\,$QL.   

Using Matthiesen's rule, the effective mobility of the top surface $\mu_\mathrm{eff, t}(L)$ indluding the two scattering mechanisms is given by Eq.~\ref{eq:mu_eff_t}. An equivalent equation is found for $\mu_\mathrm{eff, b}(L)$. These expressions enter Eq.~\ref{eq:cond} to obtain the sheet conductivity
\begin{equation*}
\sigma(L) = e
 \left[  
n_\mathrm{t}(L) 
\frac{\mu_\mathrm{t}(L)\Tilde{\mu}_\mathrm{t}(L)}{\mu_\mathrm{t}(L) + \Tilde{\mu}_\mathrm{t}(L)}
 + 
n_\mathrm{b}(L)
\frac{r_\mu\mu_\mathrm{t}(L)\Tilde{\mu}_\mathrm{t}(L)}{r_\mu\mu_\mathrm{t}(L) + \Tilde{\mu}_\mathrm{t}(L)}
 \right]    .
\end{equation*}
In this equation, $\mu_\mathrm{t}(L)$ is identical to the one from the previous model. To obtain $\Tilde{\mu}_\mathrm{t}(L)$, we fit the above formula to the experimental sheet conductivity $\sigma$. Since $n_\mathrm{t}$, $n_\mathrm{b}$ and $\mu_\mathrm{t}$ depend parametrically on $E_\mathrm{F,t}$ and $E_\mathrm{F,b}$, we have varied $E_\mathrm{F,t}$, $E_\mathrm{F,b}$ together with $r_\mu$ in order to find a range of the effective mobility of the top surface $\mu_\mathrm{eff, t}(L)$. This range is shown in Fig.~\ref{fig:fig4}. 

\begin{acknowledgments}
Funded by the Deutsche Forschungsgemeinschaft (DFG, German Research Foundation) under Germany's Excellence Strategy - Cluster of Excellence Matter and Light for Quantum Computing (ML4Q) EXC 2004/1 - 390534769. F.L. acknowledges support of the Deutsche Forschungsgemeinschaft (DFG, German Research Foundation) through priority Programme SPP 2244, project \mbox{LU 2520/1-1}.
F.S.T. acknowledges support of the Deutsche Forschungsgemeinschaft through the SFB 1083, project A12. 
% through the SPP 2244, project 443416235. 

%\section*{Author contributions}
%A.L., D.G., B.V. and F.S.T. conceived the research. A.L. performed the transport experiments. A.L., V.C. and B.V. designed the transport experiments. M.S., P.S., G.M. and D.G. carried out the MBE growth of the sample. The manuscript was written by A.L., F.L., B.V. and F.S.T. All authors discussed and commented on the manuscript.
\end{acknowledgments}

%\section*{Additional Information}
%
%\subsection*{Competing interests}
%
%The authors declare no competing interests. 

%\subsection*{Data availability}
%Raw data used to produce the figures in the main text and the Supplmentary Material are available at the J\"ulich DATA public repository.

%\bibliographystyle{apsrev4-1}
\bibliographystyle{naturemag}
\bibliography{library}

\end{document}

\title{Supplementary Information for\\
Lifting the spin-momentum locking in ultra-thin topological insulator films}

\author{Arthur Leis}
\affiliation{Peter Gr\"{u}nberg Institut (PGI-3), Forschungszentrum J\"{u}lich, 52425 J\"{u}lich, Germany}
\affiliation{J\"ulich Aachen Research Alliance (JARA), Fundamentals of Future Information Technology, 52425 J\"ulich, Germany}
\affiliation{Experimentalphysik IV A, RWTH Aachen University, Otto-Blumenthal-Stra\ss{}e, 52074 Aachen, Germany}

\author{Michael Schleenvoigt}
\affiliation{J\"ulich Aachen Research Alliance (JARA), Fundamentals of Future Information Technology, 52425 J\"ulich, Germany}
\affiliation{Peter Gr\"{u}nberg Institut (PGI-9), Forschungszentrum J\"{u}lich, 52425 J\"{u}lich, Germany}

\author{Vasily Cherepanov}
\affiliation{Peter Gr\"{u}nberg Institut (PGI-3), Forschungszentrum J\"{u}lich, 52425 J\"{u}lich, Germany}
\affiliation{J\"ulich Aachen Research Alliance (JARA), Fundamentals of Future Information Technology, 52425 J\"ulich, Germany}

\author{Felix L\"upke}
\affiliation{Peter Gr\"{u}nberg Institut (PGI-3), Forschungszentrum J\"{u}lich, 52425 J\"{u}lich, Germany}
\affiliation{J\"ulich Aachen Research Alliance (JARA), Fundamentals of Future Information Technology, 52425 J\"ulich, Germany}

\author{Peter Sch\"uffelgen}
\affiliation{J\"ulich Aachen Research Alliance (JARA), Fundamentals of Future Information Technology, 52425 J\"ulich, Germany}
\affiliation{Peter Gr\"{u}nberg Institut (PGI-9), Forschungszentrum J\"{u}lich, 52425 J\"{u}lich, Germany}

\author{Gregor Mussler}
\affiliation{J\"ulich Aachen Research Alliance (JARA), Fundamentals of Future Information Technology, 52425 J\"ulich, Germany}
\affiliation{Peter Gr\"{u}nberg Institut (PGI-9), Forschungszentrum J\"{u}lich, 52425 J\"{u}lich, Germany}

\author{Detlev Gr\"utzmacher}
\affiliation{J\"ulich Aachen Research Alliance (JARA), Fundamentals of Future Information Technology, 52425 J\"ulich, Germany}
\affiliation{Peter Gr\"{u}nberg Institut (PGI-9), Forschungszentrum J\"{u}lich, 52425 J\"{u}lich, Germany}

\author{Bert Voigtl\"ander}
\email{b.voigtlaender@fz-juelich.de}
\affiliation{Peter Gr\"{u}nberg Institut (PGI-3), Forschungszentrum J\"{u}lich, 52425 J\"{u}lich, Germany}
\affiliation{J\"ulich Aachen Research Alliance (JARA), Fundamentals of Future Information Technology, 52425 J\"ulich, Germany}
\affiliation{Experimentalphysik IV A, RWTH Aachen University, Otto-Blumenthal-Stra\ss{}e, 52074 Aachen, Germany}

\author{F. Stefan Tautz}
\affiliation{Peter Gr\"{u}nberg Institut (PGI-3), Forschungszentrum J\"{u}lich, 52425 J\"{u}lich, Germany}
\affiliation{J\"ulich Aachen Research Alliance (JARA), Fundamentals of Future Information Technology, 52425 J\"ulich, Germany}
\affiliation{Experimentalphysik IV A, RWTH Aachen University, Otto-Blumenthal-Stra\ss{}e, 52074 Aachen, Germany}

\pacs{}
\maketitle

\noindent
\subsection{Suppl.~Note 1: Four-point measurements on different terraces of (Bi$_{1-x}$Sb$_x)_2$Te$_3$}

\begin{figure}[h!]
\includegraphics[width=0.9 \linewidth]{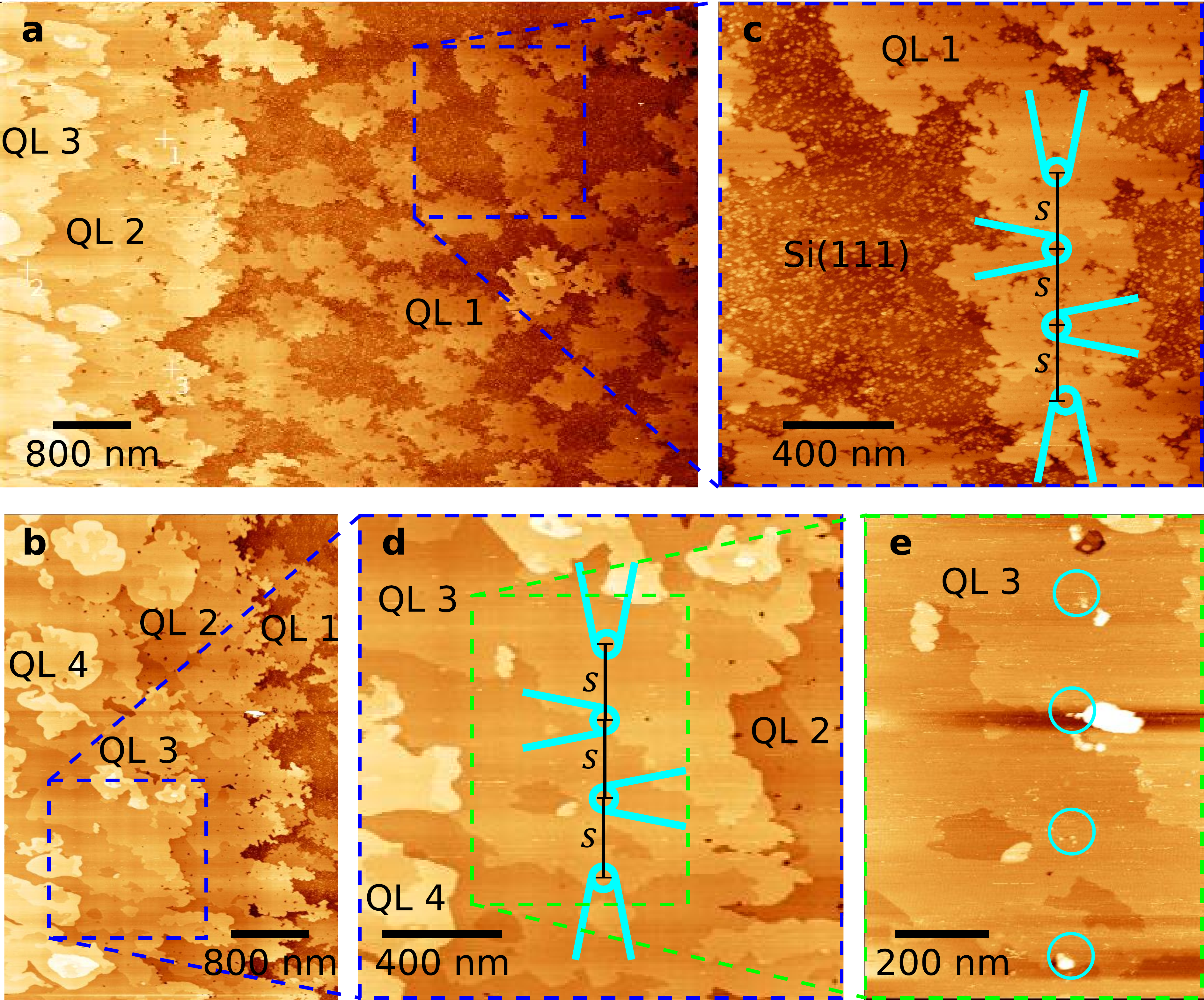}
\caption{\textbf{Four-point resistance measurements on a $1\,$QL terrace (a, b) and a $3\,$QL terrace (c, d, e).}
\textbf{a, c}, Large overview scans of the local topography serve as maps to identify terraces of different layer thicknesses.
\textbf{b, d}, Small overlapping STM scans performed with each tip are used to navigate the tips to the area of interest on the corresponding terrace. The tips (turquoise symbols) are placed in an equidistant configuration with $s \approx 250 \,$nm.
\textbf{e}, Scans of the investigated $L=3$\,QL terrace after the electrical measurements show that the contact points of the tips are discernible as small spots of only $\sim 1 \,$nm height.
Thus, hardly any damage is done to the TI film during the transport measurements.
}
\label{fig:QL1and3}
\end{figure}

\begin{figure}[h!]
\includegraphics[width=0.9\linewidth]{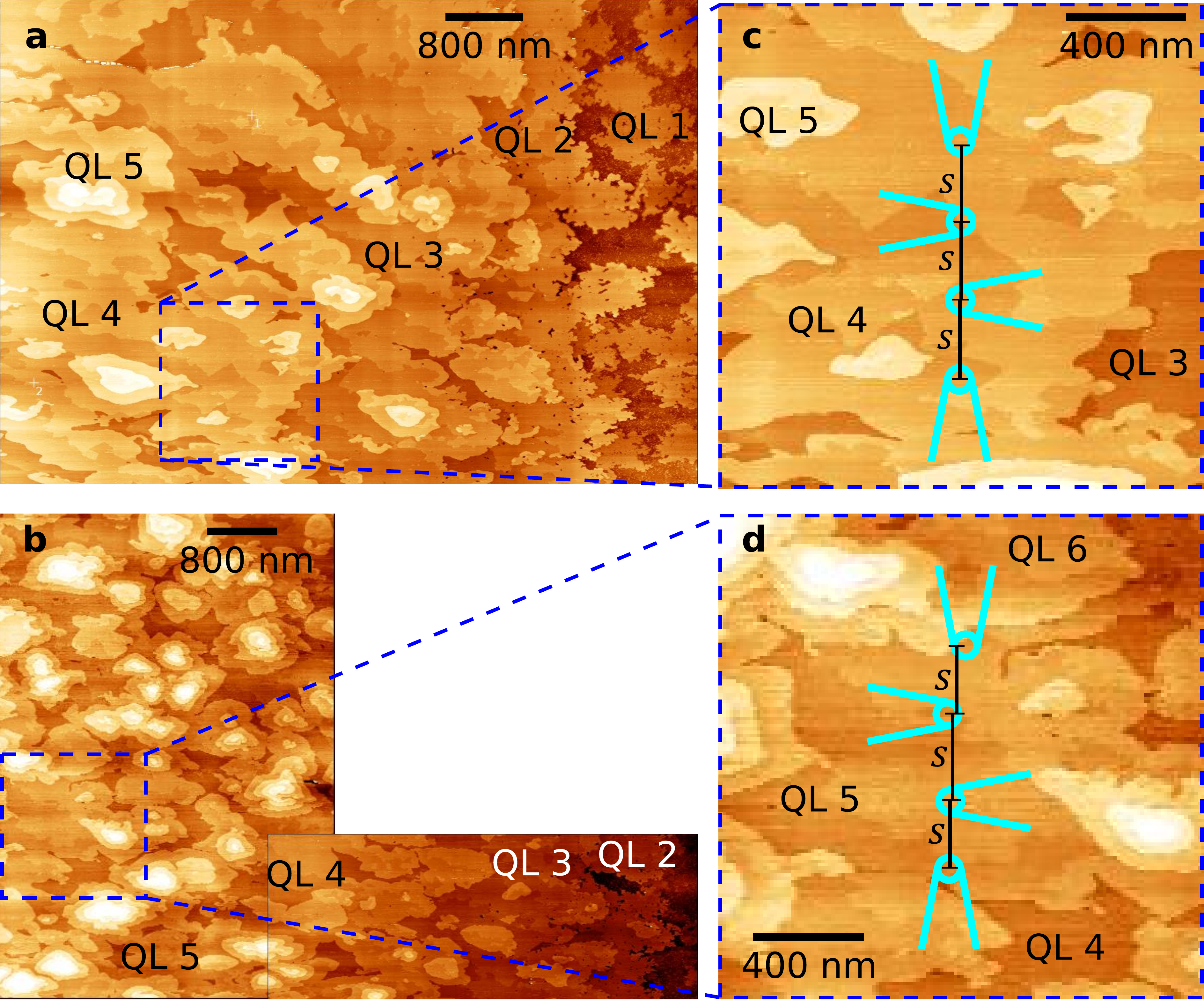}
\caption{\textbf{Four-point resistance measurement on a $4\,$QL terrace (a, b) and a $5 \,$QL terrace (c, d).}
\textbf{a, c}, Large overview scans of the local topography serve as maps to identify terraces of different layer thicknesses.
\textbf{b, d}, Small overlapping STM scans performed with each tip are used to navigate the tips to the area of interest on the corresponding terrace. The tips (turquoise symbols) are placed in an equidistant configuration with $s \approx 250 \,$nm.}
\label{fig:QL4and5}
\end{figure}

Supplementary Figures \ref{fig:QL1and3} and \ref{fig:QL4and5} show exemplary STM scans of the local topography at places where four-point resistance measurements were performed on terraces with layer thicknesses $L=1, 3, 4, 5$\,QL.

\subsection{Suppl.~Note 2: Gate-dependent transport experiments}

\begin{figure}
\includegraphics[width=\linewidth]{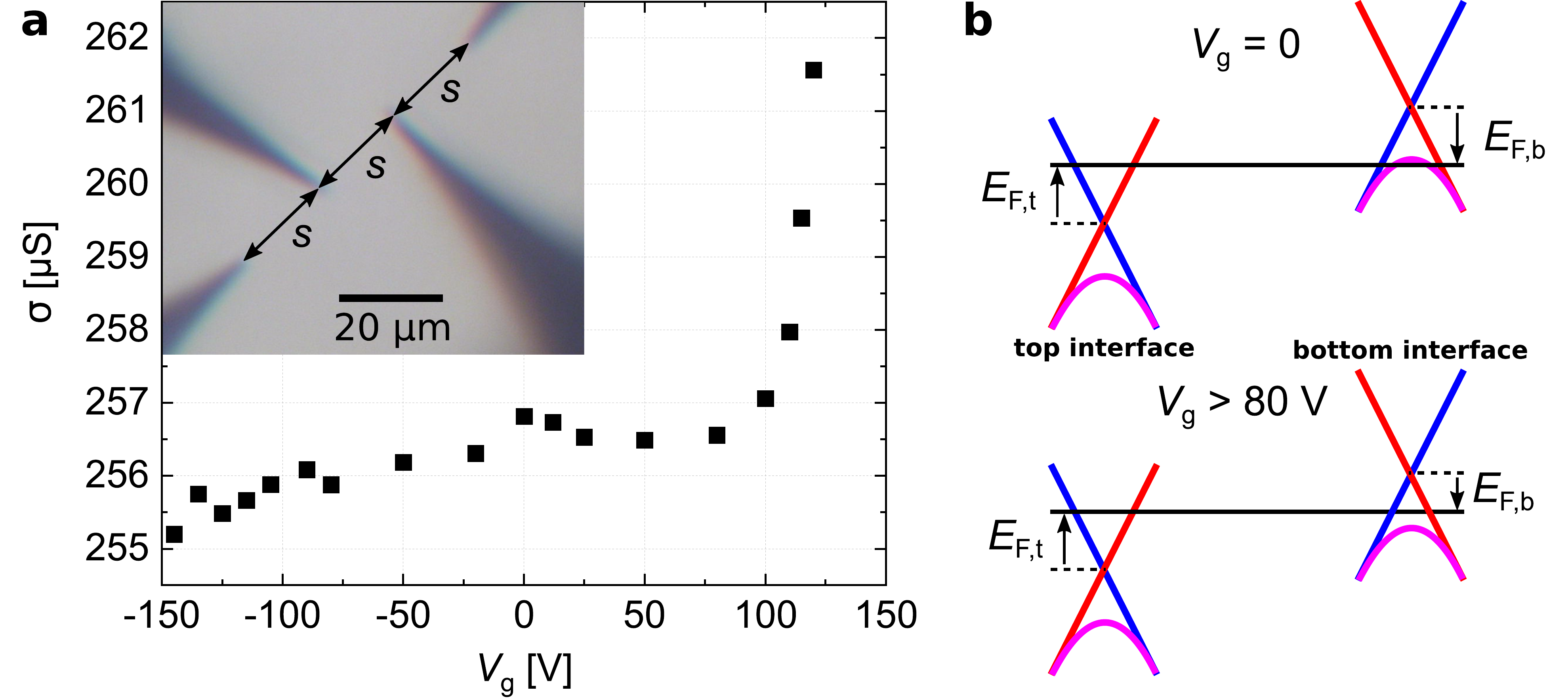}
\caption{\textbf{Gate-dependent sheet conductivity of  a $12 \,$QL (Bi$_{0.16}$Sb$_{0.84}$)$_2$Te$_3$ film.}
\textbf{a}, Sheet conductivity $\sigma$ versus back gate voltage $V_\mathrm{g}$ as obtained from four-point resistance measurements. The inset shows an optical micrograph of the corresponding equidistant tip arrangement with inter-tip distance $s \approx 20 \, \mu$m.
\textbf{b}, Schematic representation of the band structure of the TI under the influence of the gate voltage $V_\mathrm{g}$. The blue/red lines denote the linearly dispersing TSS. The bulk valence band is indicated in purple. For $V_\mathrm{g} \gtrsim 80 \,$V, the valence band at the bottom of the film is fully occupied and the Fermi level $E_\mathrm{F,b}$ is moved further upwards with increasing voltage, leading to a population of the TSS above the valence band edge.}
\label{fig:gating}
\end{figure}

Because the conductivity of the (Bi$_{1-x}$Sb$_x)_2$Te$_3$ film is observed to be insensitive to the film thickness for  $L \geq 5\,$QL, we have concluded in the paper that the two TSS at the top and bottom interfaces of the film dominate the electrical transport. While ARPES investigations yield information on the location of the Fermi level $E_\mathrm{F,t}$ on the top interface (surface) of the film, the location of the Fermi level $E_\mathrm{F,b}$ on the bottom interface between the TI film and the substrate can be deduced from gate-dependent transport experiments\textsuperscript{21} which we report in Figure \ref{fig:gating} for a 12 QL film of composition (Bi$_{0.16}$Sb$_{0.84}$)$_2$Te$_3$. Figure \ref{fig:gating}a shows the equidistant tip configuration on the sample surface with inter-tip distance $s \approx 20 \,\mu$m as well as the measured sheet conductivity as a function of the gate voltage that is applied to a back-gate contact. While there is almost no change in conductivity in the entire voltage range up to $\simeq +80 \,$V, beyond this voltage the conductivity increases sharply. 

This observation is interpreted as indicated schematically in Figure \ref{fig:gating}b. From ARPES investigations of a sample with nearly identical material composition we know that, first, the Dirac point of the TSS is located approximately $50 \,$meV above the top of the valence band\textsuperscript{25} and, second, $E_\mathrm{F,t}$ is located approximately $50 \,$meV above the Dirac point. When the gate voltage is applied, the corresponding electric field induces charge carriers in the TI, distributing them between the TSS and the bulk states at the top and the bottom interfaces. As a consequence, the bands are shifted relative to the Fermi level (Fig. \ref{fig:gating}b). However, the Fermi level $E_\mathrm{F,t}$ at the top of the film is effectively shielded from the field effect by the bottom interface and is therefore influenced only little by the back-gate\textsuperscript{21}, while the extent of the shift of $E_\mathrm{F,b}$ at the bottom interface is dependent on the magnitude of the density of states at $E_\mathrm{F,b}$.

At all gate voltages $V_\mathrm{g} \lesssim 80 \,$V, the Fermi level at the bottom interface cuts through the TSS and the bulk valence band (Fig. \ref{fig:gating}b). Consequently, the induced charge carriers mainly populate the bulk valence band, because of its in comparison to the Dirac cone much larger density of states. Since at the same time the mobility of the bulk states is comparatively low ($\simeq 2 \, \mathrm{cm}^2$/Vs) at room temperature\textsuperscript{21}, the total conductivity is almost unchanged in this gate voltage range. In contrast, at gate voltages $V_\mathrm{g}\gtrsim 80 \,$V, the bulk valence band is fully populated and $E_\mathrm{F,b}$ is shifted into the bulk band gap. Hence, the Fermi level cuts through the Dirac cone only, and the additional charge carriers that are induced by the field effect exclusively populate the TSS, resulting in a much steeper increase of the measured conductivity, because the mobility 
in the TSS is very high. 

According to the above argument, the  characteristic evolution of the gate-dependent conductivity implies that for all our measurements at $V_\mathrm{g} = 0$, the bottom Fermi level $E_\mathrm{F,b}$ of the $12\,$QL (Bi$_{0.16}$Sb$_{0.84}$)$_2$Te$_3$ film is at the valence band edge, i.e. at $E_\mathrm{F,b} \approx -50 \,$meV. The fact that $E_\mathrm{F,t} > E_\mathrm{F,b}$ is indeed in agreement with earlier studies of TI thin films and can be explained by substrate interactions\textsuperscript{21}. Moreover, the position of $E_\mathrm{F,b}$ slightly below the valence band edge for $V_\mathrm{g} = 0$ is consistent with the fact that the present sample has a composition close to pure Sb$_2$Te$_3$, which naturally exhibits hole doping, in contrast to Bi$_2$Te$_3$, which is typically electron doped.

% \section*{References} 
%\bibliographystyle{apsrev4-1}
%\bibliography{library}

%[21] -> \cite{Luepke2018}
%[25] -> \cite{Kellner2015}